\documentclass[aps,prd]{revtex4}
\usepackage{graphicx}
\usepackage{verbatim}
\usepackage{cancel}
\usepackage{physics}
\usepackage{graphicx}
\usepackage{amssymb}
\usepackage{amsmath}
\usepackage{subfig}
\usepackage{float}
\begin{document}
\title{Quantum sensing of curvature}
\author{Daniele~Bonalda, Luigi~Seveso, Matteo~G.~A.~Paris}
\affiliation{Quantum Technology Lab, Dipartimento di Fisica 
{\em Aldo Pontremoli}, Universit\`a degli Studi di Milano, 
I-20133 Milano, Italia, \email{matteo.paris@fisica.unimi.it}}
\begin{abstract}
We address the problem of sensing the curvature 
of a manifold by performing measurements on a particle 
constrained to the manifold itself. In particular, 
we consider situations where the dynamics of the particle 
is quantum mechanical and the manifold is a surface embedded 
in the three-dimensional Euclidean space. We exploit 
ideas and tools from quantum estimation theory to 
quantify the amount of information encoded into a state 
of the particle, and to seek for optimal probing schemes, 
able to actually extract this information. Explicit results
are found for a free probing particle and in the presence
of a magnetic field. We also address precision achievable 
by position measurement, and show that it provides a nearly
optimal detection scheme, at least to estimate the radius of 
a sphere or a cylinder.
\end{abstract}
\maketitle
\section{Introduction}
\addcontentsline{toc}{section}{Introduction}
In order to describe the kinematics and the dynamics 
of a physical system, from now on {\em a 
particle}, one should at first specify the ma\-nifold 
where the dynamics of the particle 
takes place, i.e.~the manifold where the 
particle propagates. Depending on the nature of the system,
this manifold may be flat or characterized by a curvature.  
In modelling a system, geometrical constraints are often 
postulated by looking at basic principles, or on the basis 
of general considerations. However, a question arises on 
whether it may be possible to obtain information about 
the manifold by a purely operational approach, i.e. by 
performing measurements on the system under investigation.
Besides the fundamental interest, sensing the curvature
has potential applications, e.g. due to the interest 
in two-dimensional curved systems, to describe physical 
effects such as Aharonov-Bohm oscillations \cite{abo}, 
formation of Landau levels \cite{l1,l2,l3,l4,l5} 
and quantum Hall effect \cite{hef}. 
\par
In this paper, we address the problem of probing  
a manifold by performing measu\-rements on a particle 
constrained to move on the manifold itself. In particular,
we focus on estimating the {\em curvature} of a manifold, and 
consider regimes where the dynamics of the particle is quantum 
mechanical. To this aim, we employ ideas and tools from quantum 
estimation theory in order to quantify the information that is
actually available according to the laws of quantum mechanics.
In addition, we look for the optimal measurement, able to extract 
the maximum information about the curvature, as well as the 
optimal preparation of the particle, i.e. the preparation which 
is most sensitive to the curvature of the underlying manifold.
\par
As a first step in this endeavour, we review the possible 
approaches to derive the Schr\"odinger equation for a particle 
constrained to a manifold, possibly subjected to an 
external field \cite{DeWitt,jen71,Costa,Costa1,Hol93,ono94,Fer,bjb13,shi16}. 
As we will see, there are at least 
two inequivalent approaches, one of which is 
more adherent to the physical situation we have in mind. We 
review the two approaches in order to establish our notation; in particular, we
discuss in some detail the differences and the similarities 
between the two methods, to illustrate the rationale 
behind our choice. 
\par
In order to quantify the available information about the 
curvature, that may be extracted by means of a measurement on
 the particle, we employ ideas and tools from quantum 
parameter estimation (QPE) theory \cite{qp1,qp12,qp2,qp3,qp4,qp5}. 
QPE generalizes to 
the quantum case the problem of point estimation arising 
in classical statistics. There, the problem is to infer
the value of a parameter by sampling from the population 
of a random variable, whose distribution depends on the 
parameter itself. An estimate of the parameter is built 
from the data sample using a point estimator, i.e. a 
parameter-space valued function of data, and the task 
is to optimise the estimation strategy with respect 
to a suitable figure of merit.  Moving to the quantum 
mechanical case leads to the introduction of a quantum statistical model, i.e. 
a family of density operators, describing the possible 
states of a quantum system. An estimate of the parameter 
is here obtained by performing a measurement and then 
processing its outcomes via a suitable estimator. The main task is 
to make the optimal choice of both the measurement scheme 
and the estimator, in order to achieve the best possible 
precision. The central figures of merit in parameter estimation are
the Fisher information (FI) and its quantum 
generalisation, the quantum Fisher information (QFI).
\par
The paper is structured as follows. In Section \ref{s:quant} we
review the possible approaches to quantization on a curved
manifold, i.e. the {\em Lagrangian} approach based on the 
use of  generalised coordinates, and the {\em Hamiltonian} 
one, where one considers the particle in $\mathbb{R}^3$, 
but forced to a two-dimensional manifold by a steep potential, 
which is constant on the surface and increases sharply 
in the normal direction. We illustrate the differences and 
the similarities between the two approaches by the specific 
examples of a particle constrained to a sphere, a cylinder, 
and a torus. In Section \ref{s:qpe} we briefly review 
classical and quantum estimation theory, introducing the 
quantum Fisher information as a measure of information about 
a parameter contained in a family of quantum states.
In Section \ref{s:free} we analyze in some details the precision 
that may be achieved in estimating the curvature of a manifold by 
a free quantum probe, i.e.~a particle that is affected only by 
the constraining potential forcing it to stay on the surface. 
In particular, we evaluate the quantum Fisher information for 
the radius of a sphere and a cylinder, and analyze its 
scaling properties with respect to time evolution and the radius 
itself. 
In Section \ref{s:field} we consider a charged quantum probe
and analyze the performances of estimation protocols in the 
presence of an external magnetic field. As we will see, the 
external field is a resource which allows one to estimate 
the radius by performing measurements on a stationary 
state, while without a field 
we need to measure the probe after a given time evolution. 
In Section \ref{s:pos} we study the Fisher Information for a 
position measurement, and show that it provides a nearly optimal
detection scheme, at least for the sphere and the 
cylinder, i.e. the Fisher Information shows the same scaling 
of the QFI, which represents its upper bound.
Section \ref{s:out} closes the paper with some concluding remarks.
\section{Schr\"odinger equation for a particle constrained 
to a manifold}\label{s:quant}
The common procedure to quantize a Hamiltonian system consists in 
introducing canonical coordinates and substituting them with self-adjoint operators, satisfying the usual commutation relations. When the underlying space is Euclidean, no significant problem is encountered. The position and momentum operators are defined (in the position 
space representation) as follows:
\begin{equation}\label{quantzation}
{Q}\psi(q)= q \psi(q) \qquad {P}\psi(q)= -i \hbar \partial_q \psi(q)\,.
\end{equation}
The only subtlety in this case is related to operator ordering ambiguities. Indeed, it is well-known that the quantum Hamiltonian is  not uniquely defined by its classical limit. That is, there  exist in general multiple Hermitian operators, i.e.~multiple functions of $Q$ and $P$, which give rise to the same phase space function when $Q$ and $P$ are replaced by $c$ numbers. When instead the underlying space is not Euclidean, but has a metric dependent on 
the coordinates, there is the additional problem that a naive quantization along the lines of (\ref{quantzation}) may lead to operators that are not self-adjoint, or do not satisfy the 
usual commutation relations. A known example is that of spherical 
coordinates, where the operator $i \hbar \partial_\theta$ is not
self-adjoint.
\par
This kind of problem may be solved in two different ways. The first 
approach is to use generalised coordinates, i.e. to assume that the 
only space existing is the manifold itself, equipped with its metric, 
and to quantize the system on the manifold itself \cite{DeWitt}. The 
second approach \cite{Costa} requires instead to consider the 
particle as living in $\mathbb{R}^3$, but forced to move only on the manifold by a steep potential, which is constant on the surface, whereas it increases sharply for every small displacement in the normal 
direction. Following this second approach, one can write the usual 
Schr\"odinger equation in the full Euclidean space, with the addition of the constraining potential term, and then separate it into two equations, corresponding to the dynamics in the normal direction and along the surface. It may be naively expected that the two methods always give rise the same dynamics; however, this is not necessarily the case. In fact, there are examples in which the two procedures lead to different Hamiltonians. 
\subsection{Quantization: Lagrangian}
Let us consider a particle of mass $M$, moving on a manifold with 
metric tensor $g$. The Christoffel symbols are defined as 
\begin{equation}
\Gamma^\rho_{\mu\sigma}= \frac{1}{2} g^{\rho \lambda}(\partial_\mu g_{\sigma\lambda} + \partial_\sigma g_{\mu\lambda}-\partial_\lambda g_{\mu\sigma})\,,
\end{equation}
and allow one to build the following Riemann tensor: 
\begin{equation}
R^\sigma_{\mu \rho \nu}=\partial_\rho \Gamma^\sigma_{\mu \nu} - \partial_\nu \Gamma^\sigma_{\mu \rho} + \Gamma^\lambda_{\mu \nu} \Gamma^\sigma_{\rho \lambda}- \Gamma^\lambda_{\mu \rho} \Gamma^\sigma_{\nu \lambda}\,.
\end{equation}
The Ricci scalar is obtained by contracting over the indexes of 
Riemann tensor as follows:
\begin{equation}
R=g^{\mu \nu} R^\lambda_{\mu \lambda \nu}\,.
\end{equation}
The classical Lagrangian of the particle is given by:
\begin{equation}\label{lagrclasselett}
L=\frac{1}{2}M g_{ij}\dot{q}^i \dot{q}^j +Q A_i g^{ij}\dot{q}_j -Q V,
\end{equation}
where $Q$ is a scalar constant, while $A_i$, $V$ are functions of the $q$'s,
which may be interpreted as the vector and the scalar potentials respectively,
describing the interaction between a particle with charge $Q$ and 
an electromagnetic field. With this interpretation Eq.~(\ref{lagrclasselett}) 
is the classical Lagrangian of a single charged particle 
interacting with an electric and magnetic field.
Upon quantizing the generalised coordinates $q_j$ the following
Schr\"odinger equation may be derived for the particle on the manifold
\cite{DeWitt} 
\begin{align}
\label{dewittfield}
i \hbar \pdv{\Psi}{t} = & -\frac{\hbar^2}{2 M \sqrt{\det g}}  \frac{\partial}{\partial q^i} (\sqrt{\det g} g^{ij} \pdv{\Psi}{q^j}) + \xi \frac{\hbar^2 }{M}R \Psi \notag \\ 
&+\frac{i Q \hbar}{2 M \sqrt{\det g}}\frac{\partial}{\partial q^i}(\sqrt{\det g} g^{ij} A_j)\Psi \\ 
& + \frac{i Q \hbar}{M}g^{ij} A_i \frac{\partial \Psi}{\partial q^j} + \frac{Q^2}{2M} g^{ij} A_i A_j \Psi + Q V \Psi\,, \notag
\end{align}
where $\xi$ is a free parameter, i.e.~we may assign to $\xi$ any real value 
and obtain the same classical theory in the limit $\hbar \rightarrow 0$.
In the case of a free particle, the above Schr\"odinger equation reduces to 
\begin{equation}\label{DeWittfree}
i \hbar \pdv{\Psi}{t} = -\frac{\hbar^2}{2 M \sqrt{\det g}}  
\frac{\partial}{\partial q^i} (\sqrt{\det g} g^{ij} \pdv{\Psi}{q^j})
+ \xi \frac{\hbar^2 }{M}R\,.
\end{equation}
The above quantization procedure is completely general and works 
for every manifold without the need to embed it in an Euclidean 
space. However, this very feature, and the presence of the free 
parameter $\xi$, pose conceptual and practical problems. On the one 
hand, it is often the case that one has to study a particle living 
on a two-dimensional surface, knowing however that it
is in reality embedded in the usual three-dimensional space. On the other hand,
with this approach one would not be able to consider the effects of
any normal field (e.g. a magnetic field normal to the surface), since
an additional dimension is unavoidably required, which is not contemplated 
by the theory. This sort of problems can be solved by an alternative 
quantization \cite{Costa,Costa1}, which we are going to review in the following 
section.
\subsection{Quantization: constraining Hamiltonian}
A more direct quantization procedure may be obtained by considering 
the particle as living in Euclidean space, but forced to stay 
on a thin layer of space around a surface by a steep potential \cite{Costa}. 
Following this approach, there is no need to quantize directly on the curved space, 
because the particle actually lives in the full Euclidean space, where 
the Schr\"odinger equation is unambiguous. Due to the nature 
of the confining potential, the Schr\"odinger equation and
the wave function may be factorized in a normal and a surface 
parts, provided that the confining potential depends on a {\em 
squeezing} parameter, and that the larger this parameter is, 
the thinner the allowed region normal to the surface. 
The Schr\"odinger equation for the surface part may be written as
\cite{Costa}
\begin{align} \label{daCostafree}
i \hbar \pdv{\Psi}{t} = & 
-\frac{\hbar^2}{2 M}\Bigg\{ \frac1{\sqrt{\det g}} \frac{\partial}{\partial q^i} \left(\sqrt{\det g} g^{ij} \pdv{\Psi}{q^j}\right) \notag \\ & + \left(\frac{1}{4}
\Tr[\alpha]^2-\det[\alpha] \right) \Psi\Bigg\} \,,
\end{align} 
where $\alpha$ is a matrix, whose elements $ \alpha_{ij}$ are the coefficients 
of the expansion of the derivatives of the normal versor on the tangent plane. 
In fact, it can be shown, and is also quite intuitive, that the derivative 
of the normal versor belongs to the tangent plane:
\begin{equation}\label{alfacoeff}
\pdv{{n}}{q^i}= \sum_{j=1}^{2} \alpha_{ij} \pdv{\vec{r}}{q^j}.
\end{equation}
It is interesting to note that these coefficients can be expressed 
as a function of the metric $g$ and of the second fundamental 
form $h$ as follows
\begin{align} \label{alphah}
\alpha_{11}&=\frac{1}{\det[g]}(g_{12} h_{21}-g_{22} h_{11})\quad
\alpha_{12}=\frac{1}{\det [g]}(g_{21} h_{11}-g_{11} h_{21})\notag\\
\alpha_{21}&=\frac{1}{\det [g]}(g_{12} h_{22}-g_{22} h_{12})\quad\alpha_{22}=\frac{1}{\det [g]}(g_{12} h_{21}-g_{11} h_{22})\,.
\end{align}
If we now introduce the notation 
\begin{align}
V_s(q_1,q_2) := - \frac{\hbar^2}{2M} \left(\frac{1}{4}
\Tr[\alpha]^2-\det[\alpha] \right)\,,
\end{align}
the quantity $V_s(q_1,q_2)$ may be interpreted 
as the surface potential due to the constraining. In terms of the mean curvature 
$C$ and the Gaussian curvature $K$ we may write
\begin{align} \label{meangaussian}
V_s(q_1,q_2)&=- \frac{\hbar^2}{2M} (C^2-K)\,,
\end{align}
where
\begin{align}
C&=\frac{1}{2 \det [g]}(g_{11}h_{22}+
g_{22}h_{11}-2g_{12}h_{12})\,, \qquad
K=\frac{\det [h]}{\det[g]}\,.
\end{align}
Besides the constraining potential, one may consider the particle 
subject to a scalar potential $V$ and a vector potential $A_i$.
Remarkably, it can be shown \cite{Fer} that no coupling appears 
between the field and the surface curvature and that, with a proper 
choice of the gauge, the surface and the transverse dynamics 
are still factorized. For the surface part, we have
\begin{align}\label{eqwithfield}
i \hbar \partial_t \Psi = &  -\frac{\hbar^2}{2 M \sqrt{\det g}} \frac{\partial}{\partial q^i} (\sqrt{\det g} g^{ij} \pdv{\Psi}{q^j})+ V_s \Psi \notag \\ &+\frac{i Q \hbar}{2 M \sqrt{\det g}}\frac{\partial}{\partial q^i}(\sqrt{\det g} g^{ij} A_j)\Psi \\ 
&+ \frac{i Q \hbar}{M}g^{ij} A_i \frac{\partial \Psi}{\partial q^j} + \frac{Q^2}{2M} g^{ij} A_i A_j \Psi + Q V \Psi \notag \,,
\end{align}
with $Q$ being the charge of the particle and $V_s$ the surface potential. 
If we compare this equation with Eq.~(\ref{dewittfield}), we realize 
that they are quite the same except for the term containing the Ricci 
scalar, that in  Eq.~\ref{eqwithfield} is replaced by the term 
containing the surface potential.
\subsubsection{Examples $\#$1: the sphere}\label{qsphere}
Let us consider the surface of a sphere of radius $\lambda$, 
parametrized in spherical coordinates, i.e. the latitude 
$\theta \in [0, \pi]$
and the longitude $\phi \in [0, 2 \pi]$. The metric matrix elements are 
given by
\begin{equation}\label{metricsphere}
g_{\theta \theta}=\lambda^2, \quad g_{\phi \phi}=
\lambda^2 \sin^2 \theta, \quad g_{\phi\theta}=g_{\theta \phi}=0.
\end{equation}
It follows that the Ricci scalar is given by $R=2/\lambda^2$ and
the surface is parametrized as $\vec{r}(\theta,\phi)=(\lambda 
\sin\theta \cos \phi, \lambda \sin \theta \sin \phi, \lambda \cos \theta)$.
The normal vector is ${n}(\theta, \phi)= \vec{r}/\lambda$, 
and it is straightforward to see that
\begin{equation}
\frac{\partial {n}}{\partial(\theta,\phi)}=\frac{1}{\lambda} 
\frac{\partial \vec{r}}{\partial(\theta,\phi)}\,.
\end{equation}
Upon comparing the above equation with Eq. (\ref{alfacoeff}), we see
that:
\begin{equation}
\alpha_{\theta \theta}= \alpha_{\phi \phi}= 1/\lambda, \quad \alpha_{\theta\phi}=\alpha_{\phi\theta}=0\,
\end{equation}
and it follows that $V_s=0$. In this case, by choosing the 
parameter $\xi$ in Eq. (\ref{dewittfield}) equal to zero, the two 
quantization procedures lead to the same Schr\"odinger equation. 
\par
For a charged particle constrained to a sphere, and subject to the effect 
of a constant magnetic field $\vec{B}$ directed along the positive $z$-axis 
($\theta=0$), we have
\begin{align}\label{sespherefield}
i \hbar \partial_t \Psi = & -\frac{\hbar^2}{2 M \lambda^2}\Big[\partial^2_\theta\Psi 
+ \cot\theta \partial_\theta \Psi +\frac{1}{\sin^2 \theta} \partial_\phi^2 \Psi \Big]
\notag \\ 
& + \frac{i Q B \hbar}{2M} \partial_\phi \Psi+ \frac{B^2 Q^2 
\lambda^2 \sin^2 \theta}{8M} \Psi\,.
\end{align}
\subsubsection{Examples $\#$2: the Cylinder} \label{cylinder}
We now consider a particle on a cylinder. The points on the 
surface have coordinates $(x=\rho \cos \theta, y=\rho \sin \theta, z)$,
where $z \in \mathbb{R}$, $\theta \in [0, 2 \pi]$, and $\rho \in [0, + \infty]$.
The surface of a cylinder of radius $\lambda$ may be parametrised by the vector 
$\vec{r}(z, \theta)=( \lambda \cos \theta, \lambda \sin \theta,z)$, and the 
metric is given by
\begin{equation}
g_{zz}=1\,, \quad g_{\theta \theta}=\lambda^2\,, \quad g_{z\theta}=g_{\theta z}=0\,.
\end{equation}
Since the metric does not depend on the coordinates ($\theta, z$), it 
immediately follows that the Ricci scalar vanishes. In order to evaluate
the $\alpha$ matrix let us consider the normal versor ${n}=(\cos \theta,
\sin \theta, 0)$, together with its derivatives:
\begin{equation}
\frac{\partial {n}}{\partial z}=0, \qquad \frac{\partial {n}}{\partial \theta}=\frac{1}{\lambda}\frac{\partial \vec{r}}{\partial \theta}\,.
\end{equation}
It follows that:
\begin{equation}
\alpha_{\theta \theta}=1/\lambda, 
\quad \alpha_{zz}=\alpha_{z \theta}= \alpha_{\theta z}=0\,.
\end{equation}
and the surface potential 
$$V_s=-\frac{\hbar^2}{8 M \lambda^2}\,,$$
does not depend on the coordinates.
Notice that in this case there is no possible choice of the 
parameter $\xi$ making the Schr\"odinger equation (\ref{dewittfield})
equal to that obtained from the constraining Hamiltonian approach, i.e.
Eq.~(\ref{eqwithfield}). This happens because the Ricci scalar is zero, 
while the surface potential is not. On the other hand, the two Hamiltonians 
differ only by the presence of the surface potential, which is a constant, 
and therefore they give rise to the same dynamics.
\par
If a magnetic field $\vec{B}$ is present, it will have a radial component,
which we denote by $\vec{B}_1$, and a normal component directed along the 
$z$-axis, which we denote by $B_0$. The radial component may be always 
chosen in order to have $\theta=0$ without loss of generality.
Such a magnetic field, in the gauge where the transverse component 
of the vector potential is zero ($A_\rho=0$), may be obtained from the 
following vector potential: $(A_\theta, A_z, A_\rho)=
(\frac{1}{2} \lambda^2 B_0, \lambda B_1 \sin \theta, 0)$. 
It follows, after some intermediate steps, that the the Schr\"odinger 
equation is given by
\begin{align}\label{schcylfield}
i \hbar \partial_t \Psi=& -\frac{\hbar^2}{2M}\Big(\frac{\partial^2_\theta \Psi}{\lambda^2} + \partial^2_z\Psi\Big)+ \frac{i\hbar Q B_0}{2M} \partial_\theta \Psi+ \notag 
\\ &+ \frac{i \hbar \lambda Q B_1}{M}\sin \theta \partial_z \Psi \notag \\ 
& + \frac{\lambda^2 Q^2}{2 M}\big( \frac{B^2_0}{4}+ B^2_1 \sin^2 \theta \big) \Psi - \frac{\hbar^2}{8 M \lambda^2}\Psi\,.
\end{align}
\subsubsection{Examples $\#$3: the Torus}
In the following, we are not investigating in details 
the case of a torus. It is however 
of interest to briefly mention it, in order to 
present a specific example where the two quantization methods 
lead to unavoidably different Hamiltonians and dynamics.
A torus is the surface generated by rotating a circle 
in three-dimensional space, about an axis coplanar with 
the circle.  As such, it is specified by two parameters: 
the radius $r$ of the circle and the distance $R$ from 
the center of the circle to the center of the torus. A 
point on the surface of a torus is identified by two 
angles: $\theta \in [0, 2 \pi]$, which tells the angle on 
the circle, and $\phi \in [0, 2 \pi]$, which measures the angle 
around the center of the torus. The corresponding 
parametrization 
reads as follows:
\begin{align}
x(\theta, \phi)=(R+r \cos \theta) \cos \phi\,, \; 
y(\theta,\phi)=(R+r \cos \theta) \sin \phi\,, \; 
z(\theta,\phi)=r \sin \theta\,,
\end{align}
and the metric is given by $g_{\theta\theta}= r^2$, 
$g_{\phi\phi}=(R+r \cos \theta)^2$, $g_{\theta\phi}=g_{\phi\theta}=0$.
The normal versor is thus $
{n}=(\cos\theta \cos \phi, \cos \theta \sin \phi, \cos \theta)^T$,
and the $\alpha$ matrix is given by
\begin{equation}
\alpha_{\theta\theta}=\frac{1}{r}\,, 
\quad \alpha_{\phi\phi}= \frac{\cos \theta}{R+r \cos \theta}\,, 
\quad \alpha_{\theta\phi}=\alpha_{\phi\theta}=0\,.
\end{equation}
The corresponding surface potential reads as follows
\begin{equation}
V_s(\theta,\phi)=-\frac{\hbar^2}{8M} \frac{R^2}{r^2 (R+r\cos\theta)^2}\,.
\end{equation}
The Ricci scalar of the torus is given by
\begin{equation}
R=\frac{2 \cos \theta }{r(R+r \cos \theta)}\,,
\end{equation}
and this means that the two procedures of quantization unavoidably 
lead to  different Hamiltonians. In fact, there is no choice of the 
parameter $\xi$ that can make the two Hamiltonians equal. Furthermore, 
the difference between the two Hamiltonians is not a constant, but 
rather it depends on the coordinates.
For a free particle, the Hamiltonian obtained with a constraining 
potential leads to following Schr\"odinger equation: 
\begin{align}
i \hbar \partial_t \Psi = & 
-\frac{\hbar^2}{2M} \bigg[ \frac{\partial^2_\theta}{r^2} - 
\frac{\sin \theta\, \partial_\theta}{r}+\frac{\partial^2_\phi}{(R+r \cos \theta)^2}
+ \bigg(\frac{R}{2r(R+r \cos \theta)}\bigg)^2 \bigg]\,\Psi\,.
\end{align}
\section{Quantum parameter estimation} \label{s:qpe}
Let us consider a family of quantum states $\rho_\lambda$, 
that are labeled by the values of a parameter of interest. 
In our case, the family consists of the possible states of 
a particle on the manifold, and the parameter corresponds to
its curvature. We refer to the particle as a {\em quantum
probe} for the parameter $\lambda$. In order to estimate 
$\lambda$, we perform the same measurement on repeated preparations
of the quantum probe, and then suitably process the sample of outcomes. 
Let us denote by $X$ the observable measured on the probe, 
and by $p(x|\lambda)$ the conditional distribution of the
outcomes, assuming that the true value of the parameter is 
$\lambda$. We also assume to perform $N$ independent 
repeated measurements on the probe. Once $X$ has been chosen 
and the sample ${\mathbf x}=\{x_1,...,x_N\}$ has been collected,
we process the data by an {\it estimator} $ \lambda \equiv 
 \lambda ({\mathbf x})$, i.e.~a function from the space 
of data to the set of possible parameter values. 
The {\it estimated value} of the parameter is the average
value of the estimator over data,~i.e.
\begin{align}
\overline{\lambda} = \int\!\! d{\mathbf x}\, p(\mathbf{x}|\lambda) 
\, \lambda ({\mathbf x})\,,
\end{align}
where $ p(\mathbf{x}|\lambda) = \prod_{k=1}^N\, p(x_k|\lambda)$, 
owing to the independence of the $M$ measurements. 
The {\it precision} of our estimation procedure is quantified by  
the variance of the estimator, i.e.: 
\begin{align}
V_\lambda \equiv \hbox{Var}\, \lambda = 
\int\!\! d{\mathbf x}\, p(\mathbf{x}|\lambda) 
\, \Big[\lambda ({\mathbf x})- \overline{\lambda}  \Big]^2\,.
\end{align}
The smaller $V_\lambda$, the more precise the 
estimation procedure. 
\par
For any (asymptotically) unbiased estimator, i.e.~any estimator satisfying
the condition $\overline{\lambda} \rightarrow \lambda$ for $N \gg1$,
there is a bound to the best achievable precision, given by the celebrated  
Cram\'er-Rao (CR) inequality:
\begin{equation}
V_\lambda \ge \frac{1}{N F_\lambda}
\end{equation}
where $F_\lambda$ is the so-called 
Fisher information (FI)
\begin{align}
F_\lambda = \int\!\! dx\, p(x|\lambda) \Big[\partial_\lambda 
\log p(x|\lambda)\Big]^2\,.
\end{align}
The most precise measurement to infer the value of 
$\lambda$ is thus the measurement maximising the 
FI, where the maximisation is performed over 
all possible observables of the probe. 
To perform such maximisation analytically, one defines the symmetric logarithmic derivative 
$L_{\lambda}$ (SLD) of the quantum statistical model, defined as the operator that satisfies the relation
\begin{equation}\label{SLD}
\frac{(L_\lambda\,\rho_\lambda+\rho_\lambda\,L_\lambda)}{2}=
\partial_\lambda \rho_\lambda\,.
\end{equation}
Then, the quantum CR theorem states that the optimal 
quantum measurements are those corresponding 
to the spectral measure of the SLD, and consequently 
$F_\lambda \leq H_\lambda = \hbox{Tr}[\rho_\lambda\,L_\lambda^2]$, 
where $H_\lambda$ is usually 
referred to as the quantum Fisher information (QFI). 
The quantum CR  inequality then states that:
\begin{equation}\label{qcr}
V_\lambda \ge \frac{1}{N H_\lambda}\,,
\end{equation}
which represents the ultimate bound to precision, i.e.~a bound taking into account both the intrinsic (quantum), 
and extrinsic (statistical), source of fluctuations 
affecting the estimator.
\par
Upon solving the eigensystem for the family of quantum states $\rho_\lambda$,
we may write $\rho_\lambda=\sum_n \rho_{n}\,\big|\phi_{n}
\rangle\langle\phi_{n}\big|$, where both the eigenvalues 
and the eigenvectors do, in general, depend on the
parameter of interest. One can then prove that the QFI can be written in the following convenient form:
\begin{equation}\label{qfidm}
H _\lambda=\sum_n\frac{(\partial_\lambda \rho_n)^2}{\rho_n}+ 
2\sum_{n\ne m}\frac{(\rho_n-\rho_m)^2}{\rho_n+\rho_m}\,\big|\langle\phi_m | 
\partial_\lambda\, \phi_{n}\rangle\big|^2\,
\end{equation}
where the sum runs over the support of $\rho_\lambda$.
The first term in Eq.~(\ref{qfidm}) is the FI of the distribution
of the eigenvalues $\rho_{n}$, whereas the second term is a positive
definite, genuinely quantum, contribution, explicitly quantifying 
the potential quantum enhancement of precision. 
When the condition $F_\lambda=H_\lambda$ is met, 
the corresponding measurement is said to be {\it optimal}. If equality is satisfied in Eq.~(\ref{qcr}), the corresponding estimator 
is said to be {\it efficient}.
\par
As it will be clear in the following, the family of states we 
are going to consider is made of pure states, ${\rho_\lambda}
=\ket{\psi_\lambda}\bra{\psi_\lambda}$, for which we have $\rho_\lambda^2={\rho_\lambda}$. From this, it follows that:
\begin{equation}
\partial_\lambda {\rho_\lambda}= \partial_\lambda \rho_\lambda^2=(\partial_\lambda {\rho_\lambda}) {\rho_\lambda}+ {\rho_\lambda}(\partial_\lambda {\rho_\lambda})\,.
\end{equation}
Comparing this equation with the definition of the Symmetric 
Logarithmic Derivative (\ref{SLD}), it immediately follows 
that for a pure quantum statistical model the SLD is given by 
\begin{equation}
{L_\lambda}= 2 \partial_\lambda {\rho}= +2 \ket{\partial_\lambda \psi}\bra{\psi}+\ket{\psi}\bra{\partial_\lambda \psi}\,.
\end{equation}
Inserting this expression into the definition of the QFI, and
using the fact that $\bra{\partial_\lambda \psi_\lambda}
\psi_\lambda\rangle$ is purely imaginary, we arrive
at 
\begin{equation} 
\label{QFIpure}
H_\lambda= 4\,\Big(\bra{\partial_\lambda \psi_\lambda}\ket{\partial_\lambda \psi_\lambda}- \left|\bra{\partial_\lambda \psi_\lambda}\ket{\psi_\lambda}\right|^2\Big)\,.
\end{equation} 
In the following, we are going to employ the above results to assess 
the performances of a quantum probe with the purpose of estimating the curvature of 
a manifold, e.g. the radius of a spherical 
or cylindrical surface. To this aim, let us denote
by $E_j$ and $\ket{\phi_j}$ the eigenvalues and the 
eigenvectors of the (time independent) particle Hamiltonian 
on the manifold, and write the generic pure state as 
$\ket{\psi_\lambda}= \sum_j c_j \ket{\phi_j}$. The state 
at time $t$ may written as 
\begin{equation}\label{pst1}
\ket{\psi_\lambda}_t= \sum_j c_j\, 
e^{-\frac{it}{\hbar}E_j}\ket{\phi_j}\,,
\end{equation}
where information about $\lambda$ is encoded in the 
eigenvalues and the eigenvectors. As such, 
we have:
\begin{equation}\label{pst2}
\ket{\partial_\lambda \psi_\lambda}_t= \sum_j c_j\, \partial_\lambda \big(e^{-\frac{it}{\hbar}E_j}\big)\ket{\phi_j}+ \sum_j c_j\,  e^{-\frac{it}{\hbar}E_j}\ket{\partial_\lambda\phi_j}\,.
\end{equation}
Upon exploiting Eqs. (\ref{pst1}) and (\ref{pst2}), we may then 
calculate $\bra{\partial_\lambda 
\psi_\lambda}\ket{\partial_\lambda 
\psi_\lambda}_t$ and $\left| \bra{\psi_\lambda}\ket{\partial_\lambda \psi_\lambda}_t 
\right|^2$ and, in turn, the QFI.  In the case of a free particle 
on a sphere,
we will explicitly calculate the QFI for a localized wave-packet, 
without needing to know its expansion on the Hamiltonian 
eigenstates. In other cases, e.g. a particle in the 
presence of a magnetic field, we will limit ourselves to 
compute the QFI for some relevant family of states, such as 
those obtained by a superposition of the ground state and a 
generic eigenstate of the system. 
\section{Sensing the curvature by a free particle}\label{s:free}
In this section, we use the tools of quantum parameter estimation
in order to assess the maximum precision that may be achieved in estimating 
the curvature of a manifold by a {\em free}  quantum particle. 
By free, we mean that the particle is affected only by the 
constraining potential that forces it to stay on the surface.  
The rationale behind this approach is that quantum systems are 
inherently sensitive to the parameters of their Hamiltonians, which may be exploited to precisely estimate the values of those
parameters by performing suitable measurements on them \cite{QCBLoss,qpl,QPCE,qpFGN,egn,meqp,fpr,wep,CVcutoff,DVcutoff}. 
This idea has been recently exploited to develop 
a new approach to probe macroscopic systems, based on the 
quantification and optimisation of the information that can 
be extracted by an interacting quantum probe as opposed to a 
classical one. In particular, in this Section, we evaluate 
the Quantum Fisher 
Information of Eq.~(\ref{QFIpure}) for the radius of a sphere, either 
using a generic pure state or an initially localized wave packet as quantum probes. We also discuss the optimization of the probe's initial preparation. We then follow the same procedure for the case of a cylinder. 
\subsection[Free particle on a sphere]{Free particle on a sphere}\label{secfreesphere}
The free Hamiltonian for a particle on a sphere of radius $\lambda$ 
may be written as 
\begin{align}\label{Spherefree}
{\cal H}_s & = 
-\frac{\hbar^2}{2 M \lambda^2 }( \cot{\theta} \partial_{\theta}+ \partial^{2}_{\theta}+ \frac{ \partial^2_\phi}{\sin^2{\theta}})
\\ & = \frac{J^2}{2 M \lambda^2}\,,
\end{align} 
where $J$ denotes the angular momentum operator. The time independent
Schr\"o\-dinger equation ${\cal H}_s|\Psi\rangle = E |\Psi\rangle$ has 
finite and separable solutions for $$(2 E M \lambda^2)/(\hbar^2)= j(j+1)\,,$$ 
with $j \in \mathbb{N}$. The eigenfunctions 
are the spherical harmonics, i.e.
\begin{align}
\label{spharmnorm}
|\Psi_{jm}\rangle & = 
\int\!\!\!\int\! d\theta\, d\phi\, \sin\theta\, Y_{jm}(\theta,\phi)\, |\theta,\phi\rangle \\ 
Y_{jm}(\theta,\phi) & = (-1)^{\frac{|m|-m}{2}}\, \sqrt{\frac{2j+1}{4 \pi}\frac{(j-|m|)!}{(j+|m|)!}} \, P^m_j(\cos{\theta})\, e^{i m \phi},
\end{align}
where $P^m_j(x)$ are the Legendre polynomials with 
$m \in \mathbb{Z}$ and $-j \leq m \leq j$.
The ket $|\theta,\phi\rangle$ in Eq. (\ref{spharmnorm}) denotes a 
localised state on the sphere, i.e.:
\begin{align}
\langle \phi^\prime,\theta^\prime|\theta,\phi\rangle & = \delta(\phi^\prime-\phi) \delta (\cos\theta^\prime-\cos\theta) = \frac{1}{|\sin\theta|} \delta(\phi^\prime-\phi) \delta (\theta^\prime-\theta)\,, \\ \langle \phi,\theta|\Psi_{jm}\rangle & = Y_{jm}(\theta,\phi)\,.
\end{align}
The corresponding eigenvalues are given by 
\begin{equation} \label{elm}
E_{jm}=\frac{\hbar^2\, j (j+1)}{2 M \lambda^2}\,.
\end{equation}
They do not depend on $m$ and are $(2j+1)-$degenerate.
We notice that while the eigenvalues do depend on the
curvature of the manifold, i.e. on the radius of the sphere,
the eigenvectors do not. This means that preparing the particle
in any given energy eigenstate and then performing a measurement cannot provide
any information about the curvature. In order to see this more explicitly,
let us remind that the spherical harmonics provide an orthonormal
basis on the sphere. Thus, we may expand any state as 
\begin{align}
|\psi\rangle & = 
\sum_{j=0}^{\infty}  \sum_{m=-j}^j c_{jm}\, 
|\Psi_{jm}\rangle\,,  \label{defjm}
\end{align}
where the amplitudes  $c_{jm}$ do not depend on $\lambda$ 
for any initial preparation of the probe particle.
At the generic time $t$, the evolved state is given by 
\begin{equation}\label{genstsph}
|\psi_\lambda\rangle  = \sum_{j=0}^{\infty} \sum_{m=-j}^j c_{jm}\, 
\, e^{-\frac{i t}{\hbar} E_{jm}}\,|\Psi_{jm}\rangle\ \,.
\end{equation}
We are now in the position of using Eq. (\ref{QFIpure}) to
evaluate the quantum Fisher information for the parameter 
$\lambda$, encoded in the generic evolved state 
$|\psi_\lambda\rangle$. In order to calculate the two 
terms involved in the QFI, we need to compute the derivative 
of the state. Using the shorthand $\sum_{mj} \equiv \sum_{j=0}^{\infty} \sum_{m=-j}^j$, we have  
\begin{align}
 |\partial_\lambda\psi_\lambda\rangle & = 
  \sum_{mj} c_{jm} e^{-\frac{i t}{\hbar} E_{jm}} 
  \left[ - i\, \frac{t}{\hbar}\, \left(\partial_\lambda E_{jm} \right)\right]
    |\Psi_{jm}\rangle  \notag \\ 
 &  =
 \sum_{mj}  \, c_{jm}\, e^{-\frac{i t}{\hbar} E_{jm}} \Big[\frac{i t \hbar\, j (j+1)}{M \lambda^3}\Big]\, |\Psi_{jm}\rangle\,,
\end{align}
where we have used the relation 
\begin{align}
\partial_\lambda E_{jm} = -\frac2\lambda\, E_{jm}\,.
\end{align}
Using the above equations, we arrive at
\begin{align}
\bra{\partial_\lambda \psi_\lambda}\ket{\partial_\lambda\psi_\lambda} 
& =  \frac{t^2 \hbar^2}{M^2 \lambda^6}
\sum_{mj}  |c_{jm}|^2  j^2 (j+1)^2\,,
\end{align}
where we used the normalisation condition for the spherical harmonics. Following similar steps, we also have:
\begin{align}
\bra{\psi_\lambda}\ket{\partial_\lambda\psi_\lambda} & =  
 i \frac{t\hbar}{M \lambda^3} 
\sum_{mj} |c_{jm}|^2 j (j+1)\,, \\
\left|\bra{\psi_\lambda}\ket{\partial_\lambda\psi_\lambda}\right|^2
& =\frac{t^2 \hbar^2}{M^2 \lambda^6} \left(
\sum_{mj} |c_{lm}|^2 j (j+1)\right)^2\,.
\end{align}
With the help of the preceding relations, we may now write down two expressions for the QFI, either in terms of the fluctuations of the Hamiltonian, or of the squared angular momentum:
\begin{align}\label{qfivariance1}
H_\lambda & =   4\, \frac{t^2 \hbar^2}{M^2 \lambda^6} \left[
\sum_{mj}  |c_{jm}|^2  j^2 (j+1)^2 - \left(
\sum_{mj} |c_{lm}|^2 j (j+1)\right)^2
\right] \\
& = \frac{16\, t^2}{\lambda^2 \hbar^2}
\left(
\left\langle {\cal H}_s^2 \right\rangle - 
\left\langle {\cal H}_s \right\rangle^2
\right) = 
\frac{16\, t^2}{\lambda^2 \hbar^2}
\left\langle \Delta{\cal H}_s^2 \right\rangle \,, \\ 
& =   \frac{4\, t^2}{M^2 \lambda^6 \hbar^2} \left(
\left\langle J^4 \right\rangle - 
\left\langle J^2 \right\rangle^2
\right) =   \frac{4\, t^2}{M^2 \lambda^6 \hbar^2} 
\left\langle \left(\Delta J^2\right)^2 \right\rangle \,,
\label{qfivariance2}
\end{align}
where, for a generic operator $O$, $\langle O \rangle$ stands for  $\langle\psi_\lambda| O | 
\psi_\lambda\rangle$.
Upon recalling the quantum Cram\'er-Rao inequality, 
we can conclude that longer time evolutions lead to a quadratic improvement in the achievable precision. This agrees with physical  intuition, since time
evolution leads to a spreading of the wave-function, i.e. it makes 
the particle {\em feel more} the effects of the curvature. 
Notice that time $t=0$ the quantum Fisher information vanishes, i.e. a 
static measurement is completely uninformative in order to 
estimate the radius. Eq. (\ref{qfivariance1}) also shows that
the smaller the radius, the more precise the estimation 
procedure, as one may have intuitively expected. Notice that 
the dependence on $\lambda$ is quite strong, i.e. the achievable precision
increases quickly as the radius decreases. At the same time, 
the dependence of the QFI on the energy fluctuations tells us 
that, in order to optimally probe the curvature, we need the 
particle to be prepared in a superposition of energy eigenstates, 
i.e. that quantum effects represent a resource for the task of 
estimating the radius.
\par
Let us now look for some specific particle initial preparations, in 
order to precisely estimate the radius, assuming that the 
overall mean energy $\bar{E}$ of the particle is fixed. 
In order to maximise the QFI, see Eq. (\ref{qfivariance1}), 
optimal states should exhibit a broad spread in energy. 
We therefore expect them to be superpositions of different energy 
eigenstates with eigenvalues as far as 
possible from $\bar{E}$. Since $\bar{E}$ is the mean 
energy, we need superpositions of some 
eigenstates with energy larger than $\bar{E}$, and some 
with smaller energy. For the sake of simplicity, and since the energy 
spectrum is bounded from below, we consider here only superpositions 
of two energy eigenstates, one of which is the ground 
state of the system, and seek for the second eigenstate in order
to maximise the QFI. Candidate optimal states are thus of the 
form $$|\varphi_{jm}\rangle = \cos\alpha\, |\Psi_{00}\rangle + 
\sin\alpha\, e^{i \beta}\, |\Psi_{jm}\rangle\,,$$ 
where the eigenstates of the free Hamiltonian 
$|\Psi_{jm}\rangle$ have been defined in Eq. 
(\ref{defjm}), and $\alpha, \beta \in [0,2\pi)$ are the 
coefficients of the superposition. Since $E_{00}=0$ we 
have 
$\bar{E}= E_{jm}\sin^2\alpha$ and
\begin{align}
\left\langle\varphi_{jm}\big| \Delta{\cal H}_s^2 \big| 
\varphi_{jm}\right\rangle = E_{jm}^2 \,\sin^2\!\!\alpha
\,\cos^2\!\!\alpha\,,
\end{align}
and, therefore, 
\begin{align}
H_\lambda & = \frac{16\,t^2}{\hbar^2\lambda^2}\, \bar{E}^2\, \frac1{\tan^2\alpha} \\
& = \frac{4\,\hbar^2\,t^2}{M^2\lambda^6}\, j^2(j+1)^2\,
\sin^{2}\!\!\alpha\,\cos^2\!\!\alpha \,. \label{sece}
\end{align}
The above expression for the QFI 
indicates that the best probing preparation is a superposition
of the ground state and an energy eigenstate with the highest possible energy 
(angular momentum), compatible with experimental constraints. For fixed 
value of $j$, the optimal superposition is a balanced one, i.e. 
$\alpha=\pi/4$, whereas for fixed value of the mean energy, 
the optimal superposition is a strongly unbalanced one, i.e. 
a state $|\varphi_{jm}\rangle$ with $j \gg1$ and small 
$\alpha\ll 1$ (while keeping fixed $\bar{E}\simeq E_{jm}\,\alpha^2$).
\par
Overall, the above requirements, are quite challenging from the 
practical point of view. Thus, we now concentrate our attention on
a more realistic example, i.e. that of an initially localized 
wave packet on the sphere, which is then left free to evolve. Assuming the particle is initially localized in a generic point on the sphere, the state reads as follows:
\begin{align}
|\psi_\kappa\rangle  = \int\!\!\!\int\! d\theta\, d\phi\, \sin\theta\, \psi_\kappa(\theta)\, |\theta,\phi\rangle\,,
\end{align}  
where the wave-function $\psi_\kappa(\theta)$ is given
in terms of the Von Mises distribution on the sphere, i.e.
\begin{equation}
\label{vonmisesstates}
\psi_\kappa(\theta)=\frac1{\sqrt{4\pi}}
\sqrt{\frac{\kappa}{\sinh \kappa}}\, e^{\frac{\kappa}{2} \cos\theta}\,.
\end{equation}
The quantity $\kappa$ is a concentration parameter. The larger $k$,
the more concentrated the distribution about the origin. For 
vanishing $\kappa$ one obtains the uniform distribution on the 
sphere. The mean energy of this state is given by
\begin{equation}
\bar{E}_\kappa=
\left\langle\psi_{\kappa}\big| {\cal H} \big| 
\psi_{\kappa}\right\rangle =\frac{\hbar^2}{4 M \lambda^2}(\kappa \coth{\kappa}-1)\,,
\end{equation}
which increases linearly with $\kappa$. Energy fluctuations are 
instead given by
\begin{equation}
\left\langle\psi_{\kappa}\big| \Delta{\cal H}^2 \big| 
\psi_{\kappa}\right\rangle =\frac{\hbar^4}{16 M^2 \lambda^4}
\left[1+ \kappa^2 \left(2- \coth^2 \kappa\right)\right]\,.
\end{equation}
leading to the following expressions for the QFI
\begin{align}
H_\lambda & =\frac{t^2\hbar^2}{ M^2 \lambda^6}
\left[1+ \kappa^2 \left(2- \coth^2 \kappa\right)\right]\,, \\ 
& = \frac{16 t^2}{\hbar^2 \lambda^2}\, \bar{E}_\kappa^2 \, g(\kappa)\,,
 \label{sec} \\ 
g(\kappa) & = \frac{1+ \kappa^2 \left(2- \coth^2 \kappa\right)}{(1-\kappa \coth{\kappa})^2}
\,. \end{align}
Upon comparing Eq. (\ref{sec}) with Eq. (\ref{sece}), we conclude that
localized states show the same scaling of optimal superpositions
with respect to the mean energy. 
Concerning localisation, we have
\begin{align}
g(\kappa) & \simeq 1 + \frac{2}{\kappa} \quad \kappa \gg 1 \notag \\
g(\kappa) & \simeq 1 + \frac{12}{\kappa^2} \quad \kappa \ll 1
\end{align}
which confirms the advantages of employing a quantum state 
(i.e. a coherent superposition) in order to probe the curvature. 
\subsection[Free particle on a cylinder]{Free particle on a cylinder}\label{frparoncyl}
According to the results of Section \ref{cylinder}, the 
free Hamiltonian for a particle constrained on an infinite cylinder 
of radius $\lambda$ may be written as 
\begin{align}
{\cal H}_c & = 
-\frac{\hbar^2}{2M}\Big(\frac{\partial^2_\theta}{\lambda^2} + 
\partial^2_z + \frac{1}{4 \lambda^2}\Big)\,,\\
& = 
\frac{1}{2M}\Big(
\frac{J^2_z}{\lambda^2} + P^2_z  -
\frac{\hbar^2}{4\lambda^2}\Big)\,,
\end{align}
where $\theta\in[0,2\pi)$ is the angular coordinate,  
$z\in {\mathbb R}$ is the axial coordinate and 
$J_z$, ${P}_z$ are the angular and the linear momentum operators 
along the $z$-axis, respectively. The corresponding 
Schr\"odinger equation is separable in an angular and an axial 
part. The eigenvectors of ${\cal H}_c$ are given by the common 
eigenstates of the commuting operators ${L}_z$ and ${P}_z$.
\begin{align}
\label{eigencyl1}
|\Phi_{km}\rangle & = \int\!\!\!\int\! d\theta\, dz\, \Phi_{km}(\theta,z)\, |\theta,z\rangle \\ 
\Phi_{km}(\theta,z)&=\frac{1}{ 2 \pi}\, e^{i k z} e^{i m \theta}\label{eigencyl2}
\end{align}
where $k$ is a real number, while $m$ is an integer number in order to satisfy 
the boundary condition $\Phi(\theta + 2 \pi,z)=\Phi(\theta,z)$. 
The corresponding eigenvalues are given by
\begin{equation} \label{eivcy}
E_{km}=\frac{\hbar^2}{2M} \Big(k^2 +\frac{m^2}{\lambda^2}-\frac{1}{4 \lambda^2}\Big)\,.
\end{equation}
In Eq. (\ref{eigencyl1}), $|\theta,z\rangle$ denotes localized states, satisfying the
relations: 
\begin{align}
\langle z,\theta |\Phi_{km}\rangle & = \Phi_{km}(\theta,z) \\ 
\langle z^\prime,\theta^\prime | \theta,z\rangle & = \delta(z^\prime-z)\delta(\theta^\prime - \theta)
\end{align}
A generic preparation of a particle of the cylinder may be thus written
as 
\begin{align}
|\psi\rangle = \sum_m\,
\int\! dk\,  c_m(k)\, |\Phi_{km}\rangle\,.
\end{align}
which evolves in time, acquiring a dependence on $\lambda$, 
according to:
\begin{align}\label{gencyl}
|\psi_\lambda\rangle = 
\sum_m\,\int\! dk\,  
c_m(k)\, e^{-\frac{it}{\hbar} E_{km}} |\Phi_{km}\rangle\,.
\end{align}
In order to calculate the QFI, we need the derivative of 
the state with respect to the cylinder radius $\lambda$.  
From Eq. (\ref{eivcy}), we have
\begin{align}
  \partial_\lambda E_{km}  = - \frac{\hbar^2}{M \lambda^3}\, 
  \left(m^2-\frac14\right)\,,
\end{align}
and therefore 
\begin{equation}
|\partial_\lambda\psi_\lambda\rangle  = 
\sum_m\,\int\! dk\,
c_m(k)\, e^{-\frac{it }{\hbar}E_{km}} \,
\left[\frac{it \hbar }{M\lambda^3}\left(m^2-\frac{1}{4}\right)\right]
|\Phi_{km}\rangle\,.
\end{equation}
Following the same procedure as for the sphere, we arrive at:
\begin{align}
\bra{\partial_\lambda \psi_\lambda}\ket{\partial_\lambda \psi_\lambda}
& = \frac{t^2 \hbar^2 }{M^2\lambda^6}
\sum_m\,\int\! dk\, |c_m(k)|^2 \left(m^2-\frac{1}{4}\right)^2 \,, \\
\bra{\psi_\lambda}\ket{\partial_\lambda \psi_\lambda}
& = i \frac{t \hbar} {M\lambda^3}
\sum_m\,\int\! dk\, |c_m(k)|^2 \left(m^2-\frac{1}{4}\right) 
\,,\\
\left|\bra{\psi_\lambda}\ket{\partial_\lambda \psi_\lambda}\right|^2
& = \frac{t^2 \hbar^2 }{M^2\lambda^6}\left[\,
\sum_m\,\int\! dk\, |c_m(k)|^2 \left(m^2-\frac{1}{4}\right) 
\right]^2\,.
\end{align}
The QFI is thus given by
\begin{align}
H_\lambda & = 
\frac{4\, t^2 \hbar^2 }{M^2\lambda^6}
\left\{
\sum_m\,\int\! dk\, |c_m(k)|^2 \left(m^2-\frac{1}{4}\right)^2
-
\left[\,
\sum_m\,\int\! dk\, |c_m(k)|^2 \left(m^2-\frac{1}{4}\right) 
\right]^2 \right\} \,,\notag \\
& = \frac{4\, t^2  }{M^2\lambda^6 \hbar^2} \left\langle(\Delta J^2_z)^2\right\rangle\,. \label{qficy}
\end{align}
As it is apparent from Eq. (\ref{qficy}),
the QFI does not contain any terms depending on
the linear momentum. This is intuitively correct, since
monitoring the motion of the particle along the $z$-axis cannot
provide any information about the curvature of the cylinder.
On the other hand, the QFI is proportional to the variance of 
the squared angular momentum along the $z$-axis, which can be seen 
as the variance of the rotational energy of the particle. This 
confirms the results already obtained for the sphere. In order to 
see this more explicitly let us choose a probing particle which is
not moving along the $z$-axis, i.e. $|c_m(k)|^2 = \delta(k)\, p_m$, where
$p_m$, with $\sum_m p_m=1$, is a generic distribution for the angular momentum.
With this choice, Eq. (\ref{qficy}) can be rewritten as:
\begin{align}
H_\lambda & = 
\frac{4\, t^2 \hbar^2 }{M^2\lambda^6} \left[\, \overline{m^4} - \left(\,\overline{m^2}\,\right)^2\right]\,, \quad \overline{m^s} = \sum_m m^s p_m\,.
\end{align}
For a uniform distribution $p_m=1/(2J)$, $m \in [-J,J]$, and for large $J$, 
we have 
\begin{align}
\overline{\!E} \simeq \frac{\hbar^2 J^2}{6  M^2\lambda^2}\;, \qquad
H_\lambda & \simeq \frac{16\, t^2 \hbar^2 J^4}{45\,  M^2\lambda^6}
\simeq \frac{64\, t^2}{\,5\,  \hbar^2 \lambda^2}\,\,\overline{\!E\, }^2\,.
\end{align}
For a probing particle prepared in a superposition $|\varphi_{J}\rangle = \frac1{\sqrt{2}}\left(|\Phi_{00}\rangle + |\Phi_{0J}\rangle\right)$ the scaling is the same, with 
a more favourable numerical factor:
\begin{align}
\overline{\!E} \simeq \frac{\hbar^2 J^2}{4  M^2\lambda^2}\;, \qquad
H_\lambda & = \frac{t^2 \hbar^2 J^4}{M^2\lambda^6}
\simeq \frac{16\, t^2}{\hbar^2 \lambda^2}\,\,\overline{\!E\, }^2\,.
\end{align}
\section{Sensing the curvature by a charged particle 
in a magnetic field} \label{s:field}
In this section, we discuss sensing protocols involving a 
charged quantum probe which, besides being constrained to the surface, is 
subject to an external magnetic field. As we will see, the presence
of an external field enhances precision, i.e. it represents a resource in
the estimation of curvature.
\subsection{Sphere in a magnetic field}
In Section \ref{s:quant}, we have discussed the Schr\"odinger 
equation for a particle of charge $Q$ and mass $M$, constrained 
to the surface of a sphere with radius $\lambda$ and subject 
to a magnetic field $B$ directed along the positive $z$-axis. 
The corresponding, time-independent, Hamiltonian may be written
as:  
\begin{align}
{\cal H}_{s\hbox{\tiny B}} = \underbrace{-\frac{\hbar^2}{2 M \lambda^2}\left[
\partial^2_\theta
+ \cot\theta \partial_\theta  +\frac{1}{\sin^2 \theta} \partial_\phi^2 \right]+ i \frac{Q B \hbar}{2M} \partial_\phi} + & 
\underbrace{\frac{B^2 Q^2 \lambda^2 \sin^2 \theta}{8M}}\,. \\
{\cal H}_{s\hbox{\tiny B}}^{(0)}\qquad \qquad \qquad \qquad \qquad & \qquad {\cal H}_{s\hbox{\tiny B}}^{(1)}\,,
\end{align}
where we have already emphasized the presence of two terms, the 
first one ${\cal H}_{s\hbox{\tiny B}}^{(0)}$ containing terms up to
the first power in the variable $y= QB $, and the second one, 
${\cal H}_{s\hbox{\tiny B}}^{(1)}$ which is of the order $O(y^2)$. 
Eigenvalues and eigenvectors of ${\cal H}_{s\hbox{\tiny B}}$ may be 
found perturbatively, based on the observation that the eigenvectors 
of the free Hamiltonian ${\cal H}_s$, i.e. the vectors 
$|\Psi_{jm}\rangle$ of Eq. (\ref{spharmnorm}), are also
eigenvectors of ${\cal H}_{s\hbox{\tiny B}}^{(0)}$ with eigenvalues 
\begin{equation}
E^{(0)}_{jm} = j(j+1)\, \frac{\hbar^2}{2M\lambda^2} - m\,
\frac{QB\hbar}{2M}\,. 
\end{equation}
Up to order $O(y^2)$, we have $E_{jm}=E^{(0)}_{jm}+E^{(1)}_{jm}$, where
\begin{align}
\label{mamma}
E^{(1)}_{jm}  = \langle\Psi_{jm}|{\cal H}_{s\hbox{\tiny B}}^{(1)}|\Psi_{jm}\rangle & = \lambda^2
\frac{Q^2 B^2}{8M}\! \int\!\!\!\int \!d \theta\, d\phi\,\sin \theta\,
Y^{*}_{jm}(\theta,\phi) \sin^2 \theta\, Y_{jm}(\theta,\phi)\,, 
\notag \\
& = (-1)^m\,\frac{ Q^2 B^2}{4M} \frac{m^2+j^2+j-1}{(2j+3)(2j-1)}\,\lambda^2\,,
\end{align}
and some standard identities involving 
associated Legendre polynomials have been employed. The corresponding eigenvectors 
$|\Xi_{jm}\rangle$ evaluate to:
\begin{align}
|\Xi_{jm}\rangle = \frac{1}{\sqrt{{\cal N}}}\left(|\Psi_{jm}\rangle + \sum_{\kappa\mu \neq jm}
\frac{\langle \Psi_{\kappa\mu}| {\cal H}_{s\hbox{\tiny B}}^{(1)}|\Psi_{jm}\rangle}{E^{(0)}_{jm}-E^{(0)}_{\kappa\mu}}\,
|\Psi_{\kappa\mu}\rangle\right)\,,
\end{align}
where ${\cal N}$ is a normalisation factor and
\begin{align}
E^{(0)}_{jm}-E^{(0)}_{\kappa\mu} & = \frac{\hbar}{2M} \left\{ \frac{\hbar}{\lambda^2}
\left[j(j+1)-\kappa (\kappa+1)\right] + (m-\mu) QB
\right\}\,, \\ 
\langle \Psi_{\kappa\mu}| {\cal H}_{s\hbox{\tiny B}}^{(1)}|\Psi_{jm}\rangle & = \delta_{j\kappa}\,\delta_{m\mu}\, E^{(1)}_{jm} + \lambda^2\, 
\frac{Q^2 B^2}{8M}\! \int\!\!\!\int \!d \theta\, d\phi\,\sin \theta\,
Y^{*}_{\kappa\mu}(\theta,\phi) \sin^2 \theta\, Y_{jm}(\theta,\phi)\,
\notag \\ & = 
\delta_{j\kappa}\,\delta_{m\mu}\, E^{(1)}_{jm} + \lambda^2\,
\frac{Q^2 B^2}{8M}\, A_{jm}\,\Big(\delta_{j,\kappa-2}  + \delta_{j,\kappa+2} \Big)\,,
 \\ \notag & \\
A_{jm} & = (-1)^{1+m}\sqrt{\frac{(j+m)(j-m)(j+m-1)(j-m-1)}{(2j+1)(2j-3)(2j-1)^2}}\,.
\end{align}
The perturbative ground state in the presence of a magnetic field
is thus given by: 
\begin{align}
|\Xi_{00}\rangle & =\frac1{\sqrt{1+g^2(\lambda)}} \Big(
|\Psi_{00}\rangle + g(\lambda)\, |\Psi_{20}\rangle \Big) \,,\\  
g(\lambda) & = \frac{Q^2 B^2 \lambda^4}{36 \sqrt{5} \hbar^2}\,.
\end{align}
Remarkably, this is a $\lambda$-dependent superposition of the 
unperturbed eigenvectors, and thus, at variance with the case
of a free particle, a measurement on 
the ground state of the system does provide information about the curvature of the sphere. The QFI may be evaluated using Eq. (\ref{QFIpure}). To this aim we compute the derivative of the 
ground state and the following scalar product:
\begin{align}
|\partial_\lambda \Xi_{00}\rangle & = \frac{\partial_\lambda g}
{\sqrt{1+g^2}}\, |\Psi_{20}\rangle - \left[
\frac{g\, \partial_\lambda g}{1+g^2}\right]\,|\Xi_{00}\rangle\,, \\
\langle\Xi_{00} |\partial_\lambda \Xi_{00}\rangle & = 0 \,, 
\end{align}
thus arriving at 
\begin{align}
H_\lambda = 4 \left(\frac{\partial_\lambda g}{1+g^2}\right)^2
= \frac{9 a^2 \lambda^6 }{(1+a^2 \lambda^8)^2}\,, 
\qquad a = \frac{Q^2 B^2}{36 \sqrt{5} \hbar^2}\,.
\end{align}
The QFI vanishes both when $\lambda\to 0$, since $H_\lambda \simeq 9 a^2 \lambda^6$ for $\lambda \ll 1$, and when $\lambda \to \infty$, since $H_\lambda \simeq 9 / (a^2 \lambda^{10})$ for $\lambda \gg 1$. It attains its maximum value $H_\lambda = \frac{45}{64} (135)^{\frac14}\sqrt{a}\simeq 2.4 \sqrt{a}$ 
for $\lambda =(3/ 5 a^2)^{\frac18}\simeq 0.94/a^{\frac14}$, a value that may be changed by tuning the magnetic field (or the charge of the quantum probe).
Overall, we have that the presence of the external field is a 
resource for curvature estimation. In particular, it allows to extract information even from measurements on the ground state of the system, i.e. a stationary state, without the need to measure the probe after a given 
time evolution.
\subsection{Cylinder in a radial magnetic field}
In Section 2.2.2 we have written the Schr\"oedinger equation 
for a particle of mass $M$, and electric charge $Q$, moving on 
the surface of a cylinder with radius $\lambda$ immersed in a 
magnetic field (\ref{schcylfield}). Let us now
focus on the specific case where the magnetic field has only 
a radial component $B_1$, whereas the axial component is vanishing 
$B_0=0$. The Hamiltonian of the system may be written as 
\begin{align}
{\cal H}_{c\hbox{\tiny B}} & =\underbrace{
-\frac{\hbar^2}{2M}\left[\frac1{\lambda^2} \left(\frac14 + 
\partial^2_\theta\right)+ \partial^2_z\right]}
+\underbrace{ i\, \frac{\,\lambda\hbar\,  Q B_1}{M}\sin \theta\, \partial_z } + \underbrace{
\frac{\,\lambda^2 Q^2B^2_1}{2 M} \sin^2 \theta}\,, \\
& \qquad \qquad \qquad \,\,
{\cal H}_{c\hbox{\tiny B}}^{(0)}
\qquad \qquad \qquad \quad \quad
{\cal H}_{c\hbox{\tiny B}}^{(1)}
\qquad \qquad \qquad \quad 
{\cal H}_{c\hbox{\tiny B}}^{(2)}\,.
\end{align}
We intend to find the eigenvalues and eigenvectors of ${\cal H}_{c\hbox{\tiny B}}$ perturbatively, at first order in the variable 
$y = QB_1$. To this aim, we neglect the effects of the term 
${\cal H}_{c\hbox{\tiny B}}^{(2)}$ and treat 
${\cal H}_{c\hbox{\tiny B}}^{(1)}$ as a perturbation to the
unperturbed Hamiltonian ${\cal H}_{c\hbox{\tiny B}}^{(0)}$.
The unperturbed eigenvalues $E_{km}$ and eigenvectors 
$|\Phi_{km}\rangle$ are those of the free Hamiltonian, and 
are given in Eqs. (\ref{eivcy}) and (\ref{eigencyl1}) 
respectively. The eigenvalues are not changed by the perturbation
since $\langle\Phi_{km}| {\cal H}_{c\hbox{\tiny B}}^{(1)}|\Phi_{km}\rangle=0$, whereas the perturbed eigenvectors $|\Upsilon_{km}\rangle$ 
are given by: 
\begin{align}
|\Upsilon_{km}\rangle = \frac{1}{\sqrt{{\cal N}}}\left(|\Phi_{km}\rangle + \sum_{\kappa\mu \neq km}
\frac{\langle \Phi_{\kappa\mu}| {\cal H}_{c\hbox{\tiny B}}^{(1)}|\Phi_{km}\rangle}{E_{km}-E_{\kappa\mu}}\,
|\Phi_{\kappa\mu}\rangle\right)\,,
\end{align}
where ${\cal N}$ is a normalisation factor and
\begin{align}
E_{km}-E_{\kappa\mu} & = \frac{\hbar^2}{2M} (k^2-\kappa^2) + \frac{\hbar^2}{2M\lambda^2}  (m^2-\mu^2) \\ 
\langle \Phi_{\kappa\mu}| {\cal H}_{c\hbox{\tiny B}}^{(1)}|\Phi_{km}\rangle & = k\, \frac{\,\lambda \hbar\, Q B_1}{M}\, \delta (k-\kappa)\, \left[\delta_{\mu-1,m}+\delta_{\mu+1,m}\right]\,.
\end{align}
The perturbed eigenstates in the presence of a magnetic field
are thus:
\begin{align}
|\Upsilon_{km}\rangle = 
\frac{1}{\sqrt{{\cal N}}}\left[|\Phi_{km}\rangle - \frac{2\, Q B_1 \lambda^3}{\hbar} \Big(B_{km}\, |\Phi_{k,m+1}\rangle +
B_{k,m-1}\, |\Phi_{k,m-1}\rangle \Big) \right]\,,
\end{align}
where
\begin{align}
B_{km} = \frac{k}{1+2m}\,, \qquad {\cal N} = 1+ 
\frac{4 Q^2 B_1^2 \lambda^6}{\hbar^2}\frac{2k^2 (4m^2+1)}{(2m+1)^2(2m-1)^2}\,.
\end{align}
The above equations implies that the ground state is left unperturbed
$|\Upsilon_{00}\rangle = |\Phi_{00}\rangle$, whereas the excited
states are affected by the perturbations.
In order to see the effect of the field on the achievable precision, let us evaluate the QFI for a generic
preparation $|\Upsilon_{km}\rangle$, $k\neq0$, of the particle. 
The derivative of the statistical model evaluates to:
\begin{align}
|\partial_\lambda\Upsilon_{km}\rangle = &  
- \frac{\partial_\lambda{\cal N}}{2{\cal N}\,} |\Upsilon_{km}\rangle 
- \frac{6\, Q B_1 \lambda^2}{\hbar} \Big(B_{km}\, |\Phi_{k,m+1}\rangle +
B_{k,m-1}\, |\Phi_{k,m-1}\rangle \Big)\,.
\end{align}
The scalar products that are needed to compute the QFI may be found by a routine calculation. The resulting QFI is given by
\begin{align}
H_\lambda = \frac{288\, a^2 k^2 (1 + 4 m^2)\, 
\lambda^4}{(1 - 4 m^2)^2 + 8 a^2 k^2 \,(1 + 4 m^2)\, \lambda^6}\,, \qquad a = \frac{Q B_1}{\hbar}\,,
\end{align}
which vanishes as $H_\lambda \propto \lambda^4$ for $\lambda \ll 1$ and 
as $H_\lambda \propto \lambda^{-2}$ for $\lambda \gg 1$, whereas it 
shows a maximum for an intermediate value of $\lambda$, which is a function of the external field and the particle charge. The lowest
useful excited state is $|\Upsilon_{10}\rangle$, corresponding to 
$H_\lambda = 288 a^2 \lambda^4/ (1+8 a^2 \lambda^6)$, which is 
maximized by $\lambda = (2a)^{-1/3}$, leading to $H_\lambda = 24 (2a)^{2/3}$. Overall, we have that, as in the previous case, 
the presence of an external magnetic field is a resource, since it allows 
 to acquire information on the radius even via a static measurement, i.e.~even when the system is prepared in a stationary state.
\section{Sensing the curvature by position measurements}\label{s:pos}
In the previous sections, we have evaluated under different scenarios the QFI, which, by means of the quantum
Cram\'er-Rao theorem, sets the ultimate bound imposed by quantum mechanics 
to the precision of any estimation protocol aimed at characterising
the curvature of a manifold. In this section, we turn our attention to position measurements, which represent the most natural choice to consider in realistic situations. For the sake of simplicity, and to maintain the section self-contained,  we focus our attention to the case of a free particle, and assess the performance of position measurements in the estimation of the radius of either a sphere or a cylinder. To this aim, we evaluate the Fisher information, and compare it to the corresponding quantum Fisher information. 
\par
\begin{figure}[h!]
\centering
\includegraphics[width=0.4\textwidth]{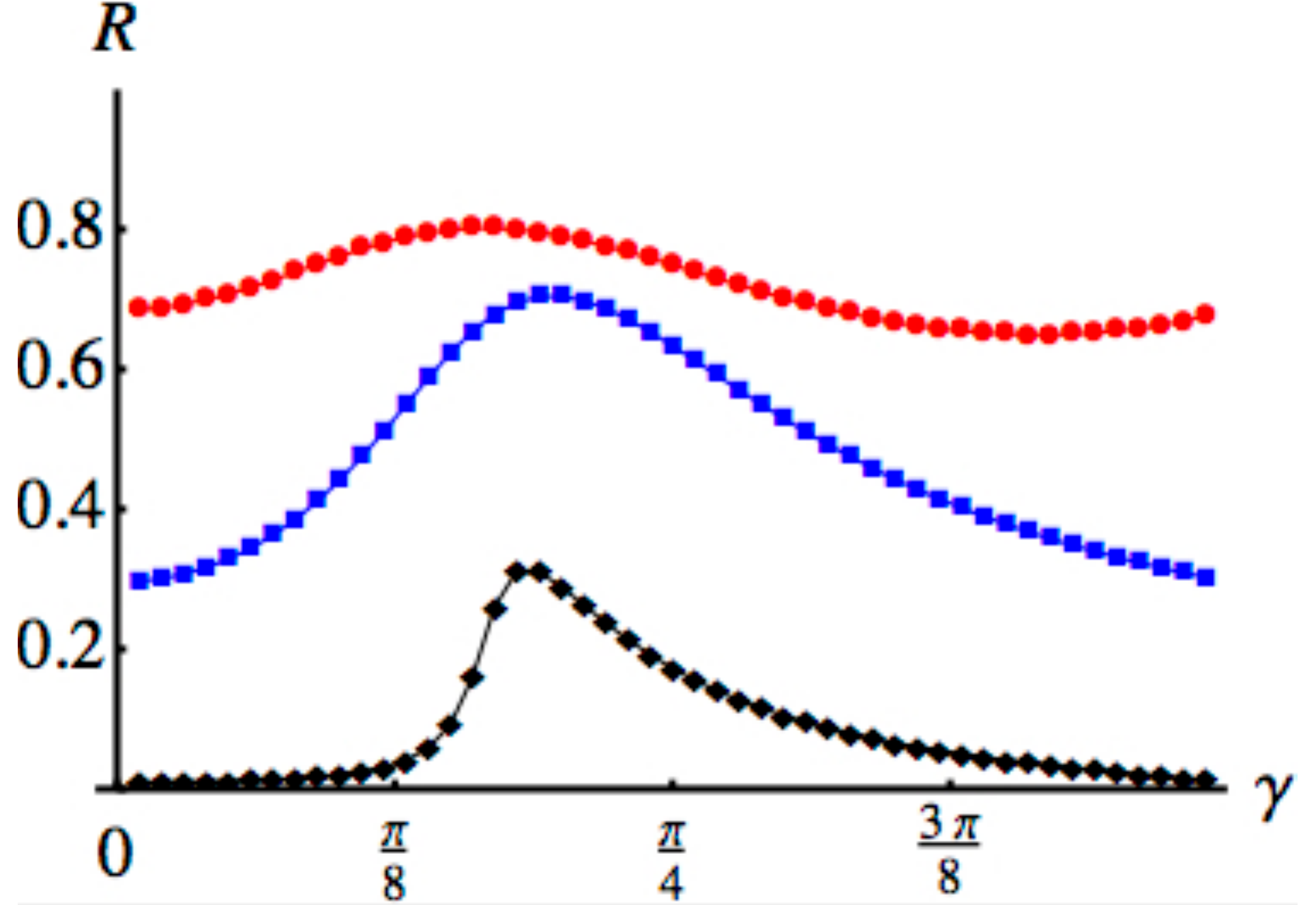}
\includegraphics[width=0.4\textwidth]{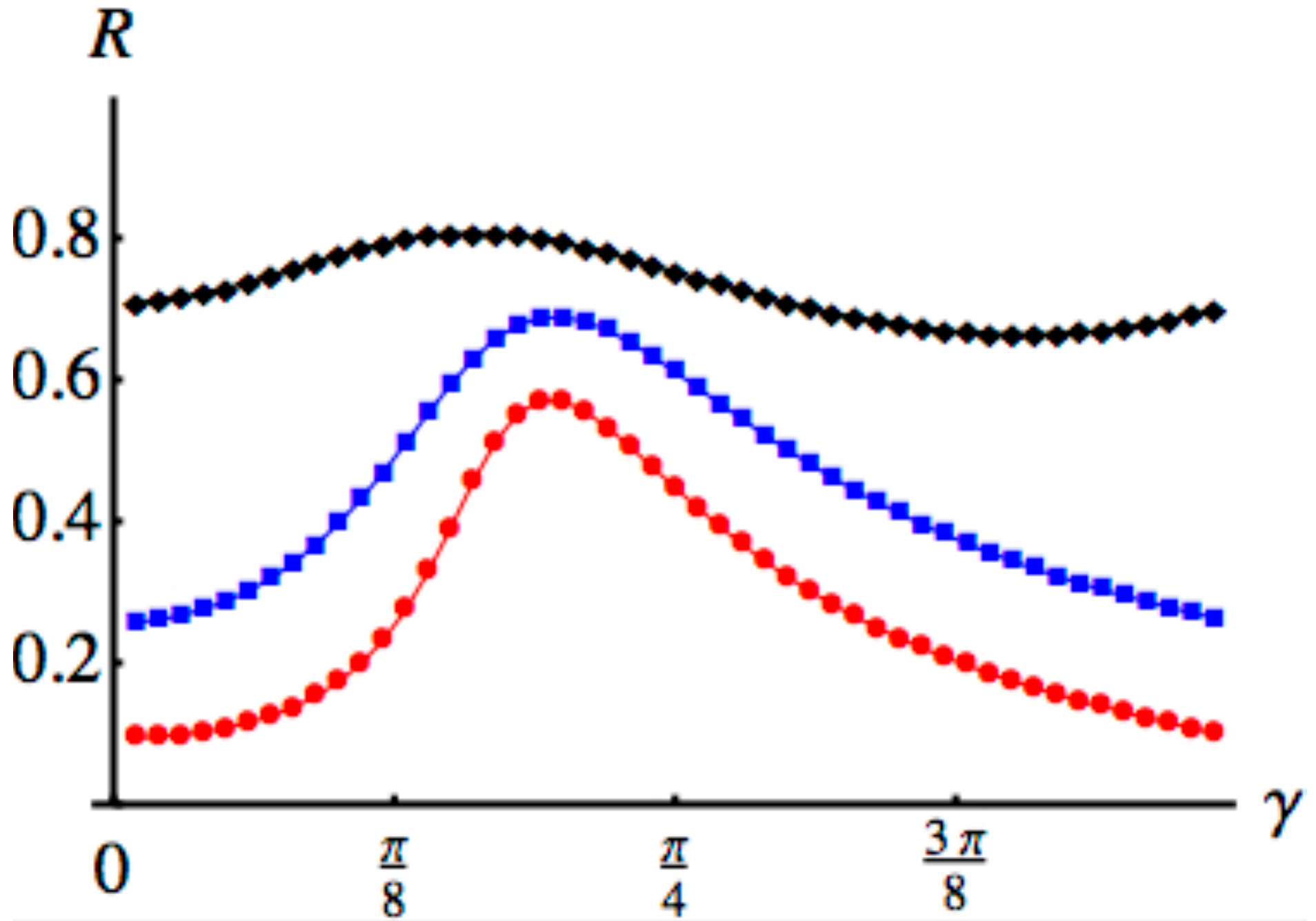}
\includegraphics[width=0.4\textwidth]{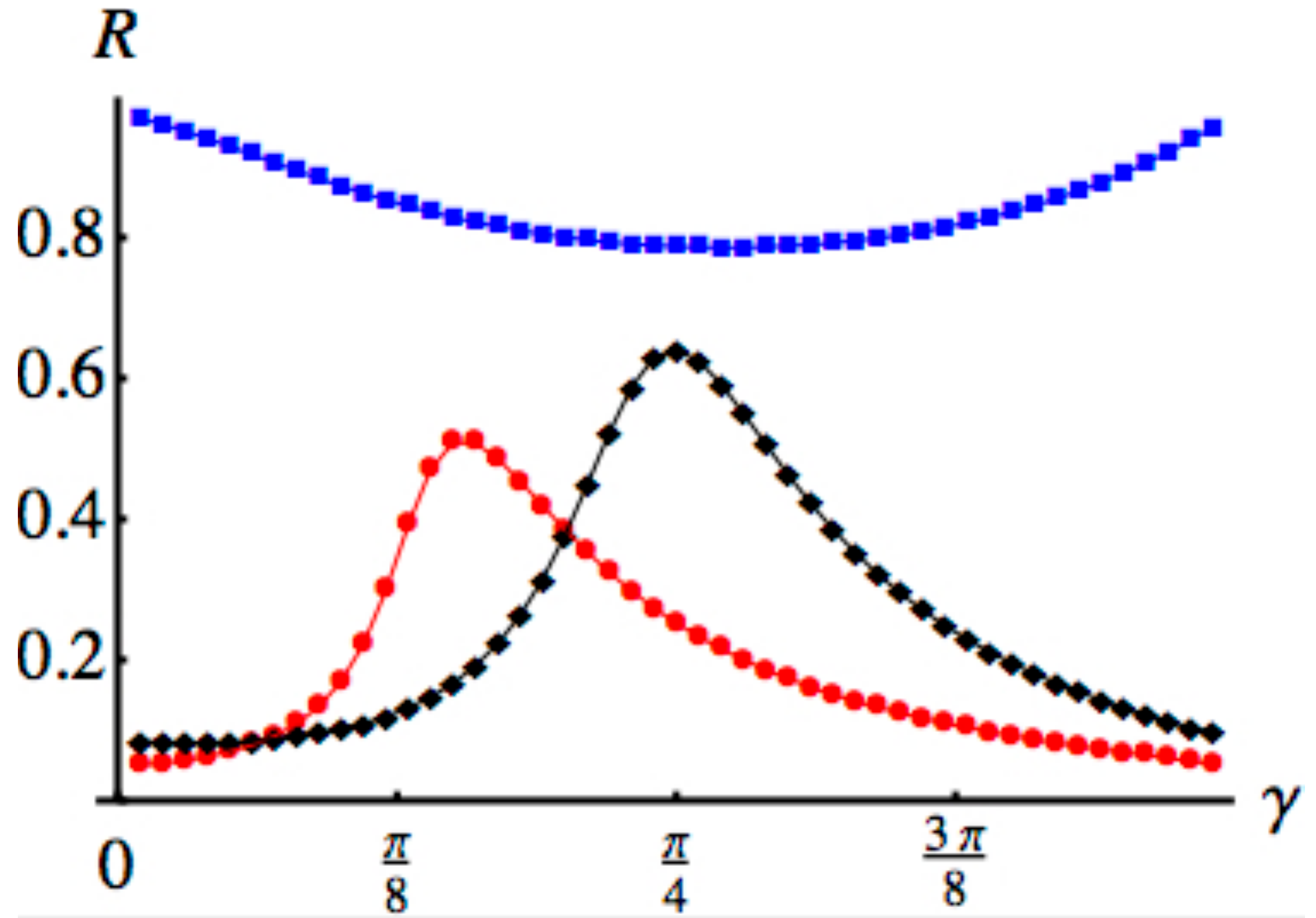}
\includegraphics[width=0.4\textwidth]{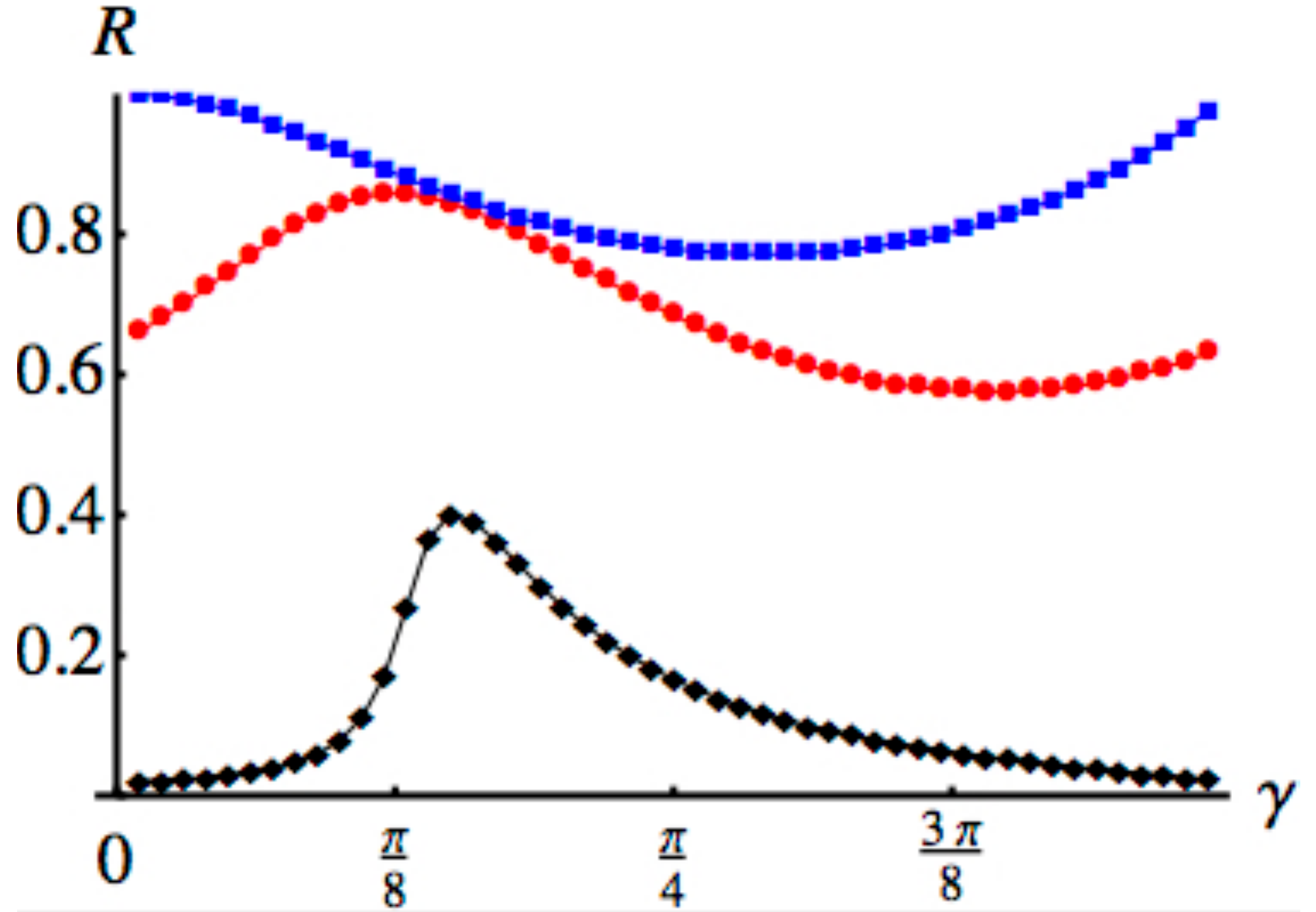}
\caption{The $R$ quantity defined in Eq. (\ref{Rq}) as a function of the superposition parameter $\gamma$. The upper panels are for $j=1$ and 
the lower ones for $j=2$.  The left panels show the behaviour of $R$ 
for $t=10$ and the right ones for $t=100$.
In each panel, red circles illustrate results for $\lambda=0.1$, 
blue squares for $\lambda=1$, and black rhombus for $\lambda=10$.
As it is apparent from the plot, there always exists a value 
of $\gamma$ for which $R$ is non negligible.} \label{f:s}
\end{figure}

The measurement of position for a particle on a sphere or a cylinder 
is described by the following set of projectors: 
\begin{align}\label{povms}
\Pi_s (\theta,\phi) & = |\theta,\phi\rangle\langle\phi,\theta|\,, 
\qquad \int_0^\pi \!\!\! d\theta\!\int_0^{2\pi}\!\!\!\!
d\phi\,\sin\theta\, \Pi_s (\theta,\phi) = {\mathbb I}\,, \\
\Pi_c (z,\phi) & = |z,\phi\rangle\langle\phi,z|\,, 
\qquad \int_{\mathbb R} \!\! dz \!\int_0^{2\pi}\!\!\!\!
d\phi\, \Pi_c (\theta,\phi) = {\mathbb I} \,.
\end{align}
Let us start by focussing our attention on the case of the sphere. Given a generic 
preparation $|\psi\rangle=\sum_{jm} c_{jm} |\Psi_{jm}\rangle$ 
of the probe, the evolved state 
$|\psi_\lambda\rangle$ is given in Eq. (\ref{genstsph}).
The position distribution at time $t$ is thus:
\begin{align}\label{p1}
p_t(\theta,\phi|\lambda) = \left|\langle\theta,\phi|
\psi_\lambda\rangle\right|^2 =  \left|\, \sum_{jm} c_{jm}\, e^{-i \frac{t}{\hbar} E_{jm}}\, Y_{jm} (\theta,\phi)\,\right|^2
\,,
\end{align}
and its derivative with respect to the radius:
\begin{align}\label{p2}
\partial_\lambda\, p_t(\theta,\phi|\lambda ) = &  
\langle\theta,\phi| 
\Big[
|\partial_\lambda\psi_\lambda\rangle\langle\psi_\lambda| + 
|\psi_\lambda\rangle\langle\partial_\lambda\psi_\lambda| 
\Big] |\theta,\phi\rangle \, \notag \\ 
= & \frac{\hbar t}{M \lambda^3} \Bigg[
i\Bigg(\sum_{jm} c_{jm}^*\, e^{i \frac{t}{\hbar} E_{jm}}\, Y_{jm}^* (\theta,\phi)\Bigg) \times \notag \\
\qquad \qquad & \times \Bigg(\sum_{jm} c_{jm}\, j(j+1)\,e^{-i \frac{t}{\hbar} E_{jm}}\, Y_{jm} (\theta,\phi)\Bigg) + c.c. \Bigg]
\end{align}
The Fisher information is given by
\begin{equation}
F_\lambda=
\int\!\!\!\int\! d\theta\, d\phi\,\sin\theta\,
\frac{\,\big[\partial_\lambda p_t(\theta,\phi|\lambda)\big]^2}{p_t(\theta,\phi|\lambda)}\,,
\end{equation}
which, using Eqs. (\ref{p1}) and (\ref{p2}), may be written as 
\begin{equation}\label{ff!}
F_\lambda=
\frac{t^2}{\lambda^6}\,\frac{\hbar^2}{M^2}\, K_s(\lambda)\,,
\end{equation}
where the function $K_s(\lambda)$ depends only weakly on $\lambda$, 
through the phase factor $e^{-i \frac{t}{\hbar} E_{jm}}$. For short
time evolution, i.e. $t \ll (2 M \lambda^2)/\hbar$, we have
$e^{-i \frac{t}{\hbar} E_{jm}} \simeq 1$ and $K$ becomes independent
of $\lambda$, i.e.:
\begin{align}
K_s \simeq 
\int\!\!\!\int\! d\theta\, d\phi\,\sin\theta\,
\frac{\,\left[i\sum_{\kappa\mu}\sum_{jm} c_{\kappa\mu}^*\, c_{jm}\, j(j+1)\, Y_{\kappa\mu}^* (\theta,\phi)
Y_{jm} (\theta,\phi) + c.c.\right]^2}{\left|\, \sum_{jm} c_{jm}\, Y_{jm} (\theta,\phi)\,\right|^2}\,.
\end{align}
Eq. (\ref{ff!}) implies that the Fisher information of position 
measurements shows the same scaling $$F_\lambda \propto t^2/\lambda^6\,,$$ 
of the QFI in Eq. (\ref{qfivariance2}), i.e. position measurements 
provide a nearly optimal detection scheme for the curvature of a 
sphere. In order to illustrate the behaviour for longer evolution times, let us 
consider the ratio between the FI in Eq. (\ref{ff!}) and the corresponding 
QFI of Eq. (\ref{qfivariance2})
\begin{align}\label{Rq}
R_t(\lambda,\gamma) = \frac{F_\lambda}{H_\lambda} = 
\frac{K_s(\lambda)}{4 \left(\Delta J^2\right)^2 }\,.
\end{align}
This quantity is bounded by the Cram\'er-Rao theorem, i.e. $0\leq R_t(\lambda,\gamma) \leq 1$,
with larger value of $R$ corresponding to situations where the performance of position 
measurements is closer to the ultimate bound, thus providing a nearly optimal detection 
scheme. In Fig. \ref{f:s} we show $R$ for a particle initially prepared in the superposition
$(\cos\gamma |\Psi_{00}\rangle + \sin\gamma|\Psi_{j0}\rangle)/\sqrt{2}$ as a function of 
$\gamma$, for different values of $t\gg1$ and different values of $j$ and $\lambda$. 
The upper panels are for $j=1$ and the lower ones for $j=2$. 
The left panels show the behaviour of $R$ for $t=10$ and the 
right ones for $t=100$.
In each panel, red circles illustrate results for $\lambda=0.1$, 
blue squares for $\lambda=1$, and black rhombi for $\lambda=10$.
As it is apparent from the plots, there always exists a value 
of $\gamma$ for which $R$ is considerably large, i.e. estimation
by position measurement is nearly optimal. We also notice that $R$ 
is not a monotone function of $t$, $j$ and $\lambda$. 
\par
For the cylinder, measurements of position are described by the operators 
$\Pi_c$ in Eq. (\ref{povms}). However, no information about the curvature 
may be obtained from measurements of the axial coordinate, and the 
relevant information is encoded instead in the distribution of the angular 
coordinate $\theta$. For a generic preparation of the probe on the 
cylinder $|\psi\rangle = \sum_m\,
\int\! dk\,  c_m(k)\, |\Phi_{km}\rangle$, the evolved state is given 
in Eq. (\ref{gencyl}), and the 
probability distribution of a $\theta$-measurement is given by 
\begin{align}
q_t(\theta|\lambda) & = \int\!\! dz \left| \langle z, \theta| \psi_\lambda\rangle\right|^2 = \frac1{2\pi} \sum_{mn} \gamma_{mn}\, e^{i(m-n)\theta} 
\, e^{- i\frac{t}{\hbar} (\epsilon_n-\epsilon_n)}\,,
\end{align}
where
\begin{align}
\epsilon_n = \frac{\hbar^2}{2 M \lambda^2}\, \left(m^2-\frac14\right)\,, \qquad
\gamma_{mn} = \int\!\! dk\, c_{m}(k) \,c^*_{n} (k)\,.
\end{align}
Therefore, we have
\begin{align}
\partial_\lambda q_t(\theta|\lambda) =  \frac{t}{\lambda^3}\frac{i\hbar}{2M\pi} \sum_{mn} \gamma_{mn} \left(m^2-n^2\right)\, e^{i(m-n)\theta} 
\, e^{- i\frac{t}{\hbar} (\epsilon_n-\epsilon_n)}\,,
\end{align}
so that the Fisher information provided by $q_t(\theta|\lambda)$ may
be written as 
\begin{equation}\label{ffc!}
F_\lambda=
\frac{t^2}{\lambda^6}\,\frac{\hbar^2}{M^2}\, K_c(\lambda)\,,
\end{equation}
where $K_c(\lambda)$ is a function only weakly dependent on 
$\lambda$, as we have already seen in the case of the sphere.
Eq. (\ref{ffc!}) shows that the FI of $q_t(\theta|\lambda)$ scales
as the corresponding QFI in Eq. (\ref{qficy}), i.e. position 
measurements (actually $\theta$-measurements) are nearly optimal 
for the purpose of detecting the curvature of a cylinder. As we have seen for the sphere,
$K_c$ becomes independent of $\lambda$ for short times. Also the 
behaviour for long times is analogous to what we have seen for the 
sphere.
\section{Conclusions}\label{s:out}
In conclusion, in this paper we have addressed the problem of 
estimating the curvature of a manifold by performing measu\-rements on a quantum particle constrained to propagate on the manifold itself. In particular,
we have focused on the case of two-dimensional 
manifolds embedded in three-dimensional Euclidean space. 
We have considered the quantum probe as living 
in the full Euclidean space, even if it is forced to remain within a thin layer of space around the surface by a steep potential. As a matter of fact, 
due to the nature of the confining potential, the Schr\"odinger 
equation and the wave function can be factorized into a normal 
and a surface components, the latter one providing a natural description
of the dynamics on the given manifold. 
\par
Upon introducing tools from quantum estimation theory, we have first 
evaluated the ultimate bound to the estimation precision for a free 
probe, i.e. a probe subject only to the constraining potential,
and have found universal scaling laws for the quantum Fisher 
information in terms of the time evolution and the radius. 
In particular, we have shown that a static measurement, i.e. 
a measurement performed right after the preparation of the 
probe, is of no use for the purpose of estimating the curvature. Rather,  the probe should be left free to evolve on the manifold in order to
acquire information about its curvature.
We have then looked at the precision bound in the presence
of an external field, showing that the field represents a 
resource, 
since it allows to exploit static measurements, e.g. on the ground 
state of the system, without the need to measure the probe after a given 
time evolution.
Finally, we have considered the performance of position measurements, proving that the corresponding
Fisher information exhibits the same scaling as the QFI with respect
to the time of evolution and the radius. Thus, position measurements provide a nearly
optimal detection scheme, at least when the unknown parameter is the radius of 
a sphere or a cylinder.
\par
Our results, in addition to their fundamental interest, pave the way
to applications based on the quantification and optimisation 
of the information, extracted via a quantum probe, about its ambient manifold.
In particular, we foresee new developments in the design of optimal 
probing strategies aimed at estimating classical geometrical parameters by means of quantum probes, thus  providing crucial ingredients for schemes of practical relevance.
\section*{Acknowledgements}
This work has been supported by SERB through project VJR/2017/000011. 
MGAP is member of GNFM-INdAM.


\begin{thebibliography}{99}
\bibitem{abo}
A. Bachtold, C. Strunk, J.-P. Salvetat, J.-M. Bonard, L. Forr\`o, 
T. Nussbaumer, and C. Sch\"onenberger, {\em Aharonov-Bohm oscillations 
in carbon nanotubes}, Nature (London) {\bf 397}, 673 (1999).
\bibitem{l1}
H. Aoki and H. Suezawa, {\em Landau quantization of electrons on a sphere}
Phys. Rev. A {\bf 46}, R1163 (1992).
\bibitem{l2}
M. Greiter, R. Thomale, {\em Landau level quantization of Dirac electrons on the sphere}, Ann. Phys. {\bf 394}, 33 (2018).
\bibitem{l3}
Ju H. Kim, I. D. Vagner, and B. Sundaram, {\em Electrons confined on the surface of a sphere in a magnetic field}, Phys. Rev. B {\bf 46}, 9501 (1992).
\bibitem{l4}
M. V. Entin and L. I. Magarill, {\em Spin-orbit interaction of electrons on a curved surface}, Phys. Rev. B {\bf 64}, 085330 (2001).
\bibitem{l5}
P. C. S. Cruz,  R. C. S.Bernardo, J. P. H. Esguerra, {\em Energy levels of a quantum particle on a cylindrical surface with non-circular cross-section in electric and magnetic fields}, Ann. Phys. {\bf 379}, 159 (2017).
\bibitem{hef}
E. Perfetto, J. Gonz\`alez, F. Guinea, S. Bellucci, and P. Onorato, 
{\em Quantum Hall effect in carbon nanotubes and curved graphene strips}, 
Phys. Rev. B {\bf 76}, 125430 (2007).
\bibitem{DeWitt} B. S. DeWitt,\emph{ Dynamical Theory in Curved Spaces. I. A Review of the Classical and Quantum Action Principles}, Rev. Mod. Phys.  \textbf{29} (1957) 377 397.
\bibitem{jen71} H. Jensen, H. Koppe, {\em Quantum mechanics with constraints}, 
Ann. Phys. {\bf 63}, 586 (1971).
\bibitem{Costa} R. C. T. da Costa,\emph{ Quantum mechanics of a costrained particle}, Phys. Rev. A  \textbf{23}, (1981) 1982.
\bibitem{Costa1} R. C. T. da Costa, \emph{Constraints in quantum mechanics}, Phys Rev A {\bf 25}, (1982) 2893
\bibitem{Hol93} P. R. Holland, {\em The Quantum Theory of Motion}, 
(Cambridge University Press) (1993).
\bibitem{ono94} C. Destri, P. Maraner, E. Onofri, {\em On the definition of 
quantum free particle on curved manifolds}, Nuovo Cim. {\bf 107}, (1994), 237.
\bibitem{Fer} G. Ferrari, G. Cuoghi, \emph{Schr\"odinger Equation for a 
Particle on a Curved Surface in an Electric and Magnetic Field}, Phys. Rev. Lett. \textbf{100}, (2008) 230403.
\bibitem{bjb13} B. J. Bernard, L C Lew Yan Voon, \emph{Notes on the quantum mechanics of particles constrained to curved surfaces}, Eur.
J. Phys. {\bf 34}, (2013), 1235.
\bibitem{shi16}
M. S. Shikakhwa, N. Chair, {\em Hamiltonian for a particle in a magnetic field on a curved surface in orthogonal curvilinear coordinates}, Phys. Lett. 
A {\bf 380}, 2876 (2016).
\bibitem{qp1}
C. W. Helstrom, {\em Cram\`er-Rao inequalities for operator-valued 
measures in quantum mechanics}, Int. J. Theor. Phys. {\bf 8}, 
361 (1973).
\bibitem{qp12}
C. W. Helstrom, {\em Estimation of a displacement parameter of a quantum system}, Int. J. Theor. Phys. {\bf 11}, 
357 (1974).
\bibitem{qp2}
A. Fujiwara and H. Nagaoka, {\em Quantum Fisher metric and 
estimation for pure state models}, Phys. Lett. A {\bf 201}, 
119 (1995).
\bibitem{qp3}
S. L. Braunstein and C. M. Caves, {\em 
Statistical distance and the geometry of quantum states},
Phys. Rev. Lett. {\bf 72}, 3439(1994).
\bibitem{qp4}
M. G. A. Paris, {\em Quantum estimation for Quantum Technology}, 
Int. J. Quantum Inf. {\bf 07}, 125 (2009).
\bibitem{qp5} L. Seveso, M. A. C. Rossi, M. G. A. Paris, \emph{Quantum metrology beyond the quantum Cram\'er-Rao theorem}, Phys. Rev. A\textbf{95} (2017) 012111.
\bibitem{QCBLoss} {Carmen Invernizzi, Matteo G. A. Paris, Stefano Pirandola}
{\em Optimal detection of losses by thermal probes}, 
Phys. Rev. A {\bf 84}, 022334 (2011).
\bibitem{qpl} A. Smirne, S. Cialdi, G. Anelli, M. G. A. Paris, B.
Vacchini, {\em Quantum probes to assess correlations in a composite 
system}, Phys. Rev. A {\bf 88}, 012108 (2013).
\bibitem{QPCE}
C. Benedetti, F. Buscemi, P. Bordone, M. G. A. Paris,
{\em Quantum probes for the spectral properties of a classical
environment},
Phys. Rev. A {\bf 89}, 032114 (2014).
\bibitem{qpFGN} M. G. A. Paris,
{\em Quantum probes for fractional Gaussian processes},
Physica A {\bf 413}, 256 (2014).
\bibitem{egn} C. Benedetti, M. G. A. Paris,
{\em Characterization of classical Gaussian processes using quantum
probes}, Phys. Lett. A {\bf 378}, 2495 (2014).
\bibitem{meqp} M. A. C. Rossi, M. G. A. Paris,
{\em Entangled quantum probes for dynamical environmental noise},
Phys. Rev. A {\bf 92}, 010302(R) (2015).
\bibitem{fpr} D. Tamascelli, S. Olivares, C. Benedetti, M. G. A. Paris, 
{\em Characterization of qubit chains by Feynman probes}, 
Phys. Rev. A {\bf 94}, 042129 (2016).
\bibitem{wep} L. Seveso, M. G. A. Paris,
{\em Can quantum probes satisfy the weak equivalence principle?},
Ann. Phys. {\bf 380}, 213 (2017).
\bibitem{CVcutoff} M. Bina, F. Grasselli, M. G. A. Paris, 
{\em Continuous-variable quantum probes for structured environments},
Phys. Rev. A \textbf{97} 012125 (2018).
\bibitem{DVcutoff} C. Benedetti, F. Salari Sehdaran, M. H. Zandi, and 
M. G. A. Paris, {\em Quantum probes for the cutoff frequency of Ohmic 
environments}, Phys. Rev. A \textbf{97} 012126 (2018).
\end{thebibliography}
\end{document}